\documentstyle[aps,prl,multicol]{revtex}

\draft
\begin{document}

\title{ Percolation of finite--sized objects on a lattice}
\author{ R. E. Amritkar$^1$ and Manojit Roy$^2$ \\
\it $^1$Physical Research Laboratory, Navrangpura, Ahmedabad 380009,
INDIA. \\
\it $^2$Department of Physics, University of Pune, Pune 411007, INDIA. }
\maketitle
\centerline{(\today)}

\begin{abstract}
We study the percolation of
{\it finite--sized} objects on two-- and three--dimensional
lattices. Our motivation stems,
on one hand from some recent interesting experimental results on
transport properties
of impurity--doped oxide perovskites \cite{OSP}, and
on the other hand from the theoretical appeal that this problem presents.
Our system exhibits a well--defined percolation threshold.
We estimate the size
of magnetic polarons, believed to be the carriers of the
abovementioned transport. We have also obtained two critical exponents for
our model, which characterize its universality class.
\end{abstract}


\begin{multicols}{2}

Over the last couple of decades percolation theory has 
generated lot of interests, both from theoretical as well as
applicational point of view \cite{SA,JWE,HK,GM,JSW,SK,AMAHK,FHS,EG,FK,MS,BK}.
Percolation is an important model exhibiting a second order phase transition
with the associated critical exponents \cite{SA,GM,JSW}.
It has also been successfully
applied to transport and phase transitions in several physical systems 
in presence of voids or impurities \cite{SA,SK,AMAHK,MS,BK}.
In this paper we investigate the percolation
mechanism of objects {\it with finite spatial extent} on a lattice.
To the best of our knowledge this problem has not been studied so far.

We are motivated for this type of study for the following reasons. Firstly,
the problem has interesting physical application. 
Recent experimental investigations in transport properties
of Fe--doped ${\rm La_{0.75}Ca_{0.25}MnO_3}$ ceramics \cite{OSP} have
thrown important light on the role of magnetic polarons which are 
finite--sized objects \cite{FR,MLS,KYT}.
In the lattice the Fe ions occupy the Mn sites. There is a 
jump in the resistivity of the system by a factor of about 80 
at about 4\% concentration of the Fe ions.
Observations of isomer shift indicate that Fe ions are in the 3+ state
only and hence cannot be expected to act as a double exchange partner
for the Mn$^{4+}$ ions. Thus the Fe impurities will be prohibiting the
transport of polarons. Using this physical picture the jump in
resistivity may be interpreted as a percolation transition for the
polarons. Secondly,
our problem has its own theoretical appeal which merits a thorough study.
It is interesting to ask whether finiteness of size
has any effect on critical exponents and hence on universality classes.
Finally, our model may have interesting application in the
transport problem
of vehicles of different sizes in a randomly grown habitation. The last
problem has an additional feature that not only the objects themselves
but also the obstacles have finite sizes. We shall see that in some
cases the two problems can be mapped into each other.

Our observations bring out a well--defined percolation threshold 
probability for our model in both two-- and three--dimension.
We have analysed our results using the ansatz of scaling 
due to the finite lattice size. We also estimate some critical
exponents near the threshold.

The model we consider is as follows. We take a lattice
and randomly disallow its sites with a probability $q$.
We define our percolating object as a spatially extended entity
of linear dimension $r$ (in lattice units) consisting of $n(r)$ sites.
We now say that such an object is allowed to percolate
in the lattice only if none of its $n(r)$ sites overlaps with any of the
disallowed sites. We then study the standard percolation problem
for such an object.

One can also study the following complementary 
problem. Let us place obstacles of linear size $r$ containing $n(r)$ sites
with a suitably defined center at random locations
with probability $q$. We now treat the remaining sites as allowed and
study the standard site percolation problem for point objects.
It is easy to see
that the two problems are essentially equivalent under the following
conditions: firstly,
the centers of the finite--sized objects or obstacles lie on a site;
secondly, the objects or obstacles
themselves have the same symmetry as that of the underlying lattice;
and finally, the obstacles are allowed to overlap \cite{AR1}.
Numerically it is
easier to study the problem with obstacles, which we use for calculations
in this paper.

We now present our main results \cite{AR2}. We consider 
a two dimensional square lattice and a three dimensional simple cubic
lattice for our study. We restrict ourselves to the cases of objects with
circular symmetry in two dimension and spherical symmetry in three dimension,
with $r$ as radius and center on a lattice site.
We have used periodic boundary conditions for
our system (otherwise finite-sized objects are ill--defined at and near
the lattice boundary). Different linear sizes $L$ of the
lattice are considered: for two dimension, we have taken $L$ = 10, 20, 40,
80, 160, 320, 640, 1280, 2560, and 5120, and for three dimension, $L$ = 10,
20, 40, 80, 160, and 250.
We have varied probability $q$ between 0 and 1, and also used
different values of the radius $r$ (first column of Table \ref{table1}
shows various $r$ values for two--dimensional
lattice), starting with the nearest neighbor
case ($r$=1 lattice unit). Data are averaged over many
realizations of randomly generated disallowed sites for each
value of $q$. It is to be noted that no exact analysis
is possible for our model in two-- or three--dimension, though
some exact results can be obtained for
one dimensional lattice and Bethe lattice \cite{AR3}.

We define a {\it cluster} as a group of sites with the following
properties: centers of the percolating objects can be placed on these
sites, and, these sites are connected through their nearest neighbors.
We call a lattice as {\it percolating} for a
given $q$ and $r$, when there exists at least one cluster
spanning the lattice end to end (a so called `infinite cluster').
We define {\it threshold probability} $q_c$ such that for $q > q_c$
the lattice ceases to be percolating. It is obvious that $q_c$ is a
function of the radius $r$.

Let $F(q)$ denote the fraction of percolating realizations (those
which support
a spanning cluster). Figures 1 and 2 show, respectively for two-- and
three--dimension, the variation of $F(q)$
with the probability $q$, for different lattice sizes $L$ and
radius $r$=1. Both plots exhibit an approach
towards a step function of the form
\begin{eqnarray}
F(q) \;\;\;& = & \;\;\;1\;\;\;\;\;\;{\rm for}\;\;\;q < q_c\;, \label{P2} \\
     \;\;\;& = & \;\;\;0\;\;\;\;\;\;{\rm for}\;\;\;q > q_c\;, \nonumber
\end{eqnarray}
as $L$ increases to $\infty$. This behavior remains essentially the
same for all other values of $r$ that we have considered.
The data of both Fig.~1 and Fig.~2 are seen to obey a
finite (lattice) size scaling relation, near the threshold $q_c$, of the type
\begin{eqnarray}
F(q,L) \;\;\;\rightarrow \;\;\;L^{-b}\;G(L^a \vert q - q_c \vert)\;, \label{P1}
\end{eqnarray}
where $a$ and $b$ are the scaling exponents.
We have seen that all the graphs of Fig.~1
do collapse onto a single function which is almost a step function
of the form (\ref{P2}), and same is true for the graphs of Fig.~2.
This confirms the existence of
scaling behavior (\ref{P1}). We have obtained the exponents
$a$, $b$ for various radii $r$ in both the dimensions.
As $r$ increases from 1 to $\sqrt{8}$, $a$ decreases monotonically from
0.8 to 0.7 for two dimensional square lattice, whereas for three
dimensional simple cubic lattice it reduces from 1.17 to 0.96; value
of $b$ remains zero for all $r$ in both cases.
We get an estimate of $q_c$ using this scaling
scheme, as $L\rightarrow \infty$. In third column of Table \ref{table1}
we list these values for all finite $r$ for the two--dimensional lattice.
There is one more way of obtaining $q_c$ as explained
below. The graphs of Fig.~1 and Fig.~2 are approximately linear for values
of $F(q)$
between 0.3 and 0.6. We consider the $q$--intercepts of these lines.
The extrapolated value of this intercept
as $1/L \rightarrow 0$ yields $q_c$. These values 
are almost identical to those listed in Table~\ref{table1},
thereby exhibiting internal consistency of our calculations.

One can calculate the probability $p$ that a given lattice site is
allowed to be the center of a percolating object.
Quite simply, the required probability is
\begin{eqnarray}
p(q,r) = (1 - q)^{n(r)}\;,  \label{P3}
\end{eqnarray}
where $q$ is our original probability and $n(r)$ is the number of lattice
sites contained in the object. Second column of Table~\ref{table1}
shows the values of $n(r)$ for all $r$, in two dimension.
We list, in fourth column, the 
values of $p(q_c,r)$ for different radii $r$ at the threshold $q_c$.

Table~\ref{table2} shows, for three dimensional simple cubic lattice,
the various $r$ values and the corresponding
quantities $n(r)$, $q_c$ and $p(q_c,r)$.
We notice that for both two and three dimensions, the objects with finite $r$
are percolating at lower $p$ values than the
point objects. The reason for this becomes clear if we look at the equivalent
problem of obstacles. Here the sites are disallowed in clumps rather than
in a homogeneous manner throughout the lattice, thereby leaving more
channels open for percolation. This effect is more pronounced the higher
the radius. For instance, in three dimension one sees that for
$r=\sqrt{6}$ the lattice percolates even with the removal of more than
90\% of its sites !. This also brings out another interesting feature.
The dependence of $p(q_c,r)$ on $r$ is not monotonous; it exhibits
occasional peaks (in Table~\ref{table1} one peak is at $r=2$, and in
Table~\ref{table2} two peaks can be seen at $r=2$ and $\sqrt{8}$).

As discussed in the introduction, the jump in the resistivity of
Fe--doped ceramic ${\rm La_{0.75}Ca_{0.25}MnO_3}$ at about 4\%
concentration of Fe ions may be interpreted as a percolation transition
for the polarons. By assuming a homogeneous and uniform distribution
of Fe ions, Ogale et. al. have suggested that the polaron radius is
about one lattice unit. However, the Fe ions are more likely to be
randomly distributed in the lattice. This leaves many channels open
for transport and thereby allowing objects with bigger radii to pass
through. Table \ref{table2} (for three dimensional
case) indicates a $q_c$ value 0.04 for radius $r$ between 2 and $\sqrt{5}$.
This suggests a polaron radius to be slightly larger than
two lattice units \cite{AR4}.

As in the standard percolation problem we find clusters of various sizes.
We have investigated the distribution of maximum size $S_m$ (normalized
by the number of lattice sites) of the
cluster, averaged over lattice realizations, for different values of
$q$ and $r$. Figure 3 plots the variation of $S_m$
with $q$, for different lattice
sizes and the radius $r$=1, in a two dimensional lattice.
The plot exhibits a sharp fall
at the threshold $q_c$ for large $L$. This behavior remains essentially 
similar for all other values of $r$. 
A similar finite--size scaling behavior as for percolating fraction $F(q)$
(Eq.~\ref{P1}) is observed for $S_m$. We have estimated the corresponding
scaling exponents $a$, $b$. $a$ varies in
essentially the same way in both the dimensions as stated before for
the case of $F(q)$. The exponent $b$, however, has 
nonzero values: $b$ = 0.14 for two dimensional lattice, whereas $b$ = 
0.5 for three dimensional lattice, independent of the radii $r$.

We have studied the nature of the decay of Fig.~3 near
$q_c$. We observe this to be a {\it power law} of the form
\begin{eqnarray}
S_m(q) \propto (q_c - q)^\beta\;.  \label{P4}
\end{eqnarray}
We have estimated the exponent $\beta$ of the relation~(\ref{P4}) for both
two-- and three--dimensional lattices with different radii $r$.
Fifth column of Table~\ref{table1} lists the $\beta$ values for
different $r$, for the two dimensional lattice. The error margin for our
estimates is $\pm 0.01$. We have also included the corresponding exponent for
point object (taken from \cite{SA}).
We have not listed our $\beta$--estimates for
three dimensional lattice because the error margin was not within
acceptable limit (the
reason is that one has to explore lattice sizes larger than
$L=250$ which is the maximum size we could study due to limited
computer resource \cite{AR2}).

We have also studied the distribution of average size $S_a$ of the clusters,
excluding the infinite cluster, as $q$ and $r$ are varied.
We define $S_a$ following Stauffer et al. \cite{SA}. The probability
that an arbitrary site belongs to any finite cluster in the lattice is
$\sum_s n_ss\;,$
where $n_s$ is the number of the clusters of size $s$, normalized
by the number of lattice sites.
Then $w_s = n_ss/\sum n_ss$ is the probability that the cluster 
to which an arbitrary allowed site belongs has a size $s$. The average
size $S_a$ is therefore
\begin{eqnarray}
S_a = \sum_s w_ss = \sum \big(n_ss^2 / \sum n_ss \big)\;.  \label{P5}
\end{eqnarray}
As explained in Ref.~\cite{SA}, we take Eq.~(\ref{P5}) as the definition
of our mean size and not the more familiar expression $\sum n_ss/\sum n_s$,
because in Eq.~(\ref{P5}) lattice sites, rather than the clusters,
are selected with equal probability.
The average size $S_a$ shows a diverging trend near $q_c$ from
both sides. It also exhibits finite--size scaling as for $F(q)$
(Eq.~(\ref{P1})) and $S_m$. We have estimated the scaling
exponents $a$, $b$ for $S_a$. We find that $a$ varies
with $r$ in the same way as in the earlier two cases of $F$ and $S_m$.
Exponent $b$ takes value $-1.64$ in two dimension, $-1.7$ in three 
dimension, irrespective of the radius $r$. 

We have investigated the nature of divergence of $S_a$ near the
threshold $q_c$. We find it to be a {\it power law} of the type
\begin{eqnarray}
S_a(q) \propto \vert q_c - q \vert^{-\gamma}\;, \label{P6}
\end{eqnarray}
where $\gamma$ is the power--law exponent. In last column of
Table~\ref{table1}
we list the values of $\gamma$ for various $r$, for two dimensional lattice.
The corresponding error margin is $\pm 0.1$.
We also show the value for point object case \cite{SA} for comparison.
We have not shown the $\gamma$ values
for three dimensional lattice, for reasons stated earlier.

The question that remains to be answered is: whether the universality
class for our model, indicated by the exponents $\beta$ and $\gamma$, is
same as, or distinct from, that of the point percolation case.
The similarity of the values of 
$\beta$ and $\gamma$, within their respective error margins,
to their $r$=0 counterparts may suggest these two to be the same.
However, possibility exists that by taking larger lattice sizes and
with consequent reduction in error one gets a quite distinct universality
class for finite--sized object case.
Unfortunately, no definitive conclusion can be drawn at this stage.

In conclusion, we have, to our knowledge for the first time,
investigated the percolation mechanism of finite--sized objects.
We observe a well--defined percolating threshold for our model,
in a two dimensional
square lattice and three dimensional simple cubic lattice,
the threshold depending on the radius of the percolating objects.
Based on this study,
we have made an estimate of the size of polarons which are
believed to be carriers of transport in oxide perovskites \cite{AR5}. In our
model there
exists a scaling due to the finite size of the lattice, thereby 
allowing us to obtain important quantities for infinite systems from
finite samples. We have also obtained two critical exponents,
which characterize the universality class for our system. We
expect that our model will be useful in characterizing the problem of transport
of finite sized objects, such as finite sized excitations in solids
and heavy vehicles in a randomly grown habitation.

\vspace{.3 in}
\noindent
One of the authors (REA) acknowledges Department of Science and
Technology (India)
and the other (MR) acknowledges University Grants Commission (India)
for financial assistance.

\newpage

\narrowtext
\vbox{
\begin{table}
\caption{The quantities $n(r)$, the number of sites in an object,
$q_c$, the threshold probability, $p(q_c,r)$, the probability for the
allowed sites for the center of an object at $q_c$,
and $\beta$ and $\gamma$, the critical exponents,
for different radii $r$ in two dimensional square lattice.
Error margins for $q_c$, $\beta$, and $\gamma$ are $\pm 0.0001$,
$\pm 0.01$, and $\pm 0.1$ respectively. Values for point object
($r=0$) are from Ref.[1].}
\begin{tabular}{cccccc}
$r$ & $n(r)$ & $q_c$ & $p(q_c,r)$ & $\beta$ & $\gamma$ \\
\tableline
0 & 1 & -- & 0.5928 & 0.14 & 2.39 \\
1 & 5 & 0.1153 & 0.5420 & 0.14 & 2.30 \\
$\sqrt{2}$ & 9 & 0.0868 & 0.4417 & 0.14 & 2.29 \\
2 & 13 & 0.0538 & 0.4873 & 0.15 & 2.22 \\
$\sqrt{5}$ & 21 & 0.0406 & 0.4188 & 0.15 & 2.28 \\
$\sqrt{8}$ & 25 & 0.0358 & 0.4020 & 0.14 & 2.39 \\
\end{tabular}
\label{table1}
\end{table}
}

\narrowtext
\vbox{
\begin{table}
\caption{The quantities $n(r), q_c, p(q_c,r)$, for different $r$
values in three dimensional simple cubic lattice. Error
margin for $q_c$ is $\pm 0.0001$. $p(q_c,r)$ for $r=0$ is
from Ref.[1].}
\begin{tabular}{cccc}
$r$ & $n(r)$ & $q_c$ & $p(q_c,r)$ \\
\tableline
0 & 1 & -- & 0.3116 \\
1 & 7 & 0.1921 & 0.2247 \\
$\sqrt{2}$ & 19 & 0.0912 & 0.1625 \\
$\sqrt{3}$ & 27 & 0.0752 & 0.1211 \\
2 & 33 & 0.0591 & 0.1340 \\
$\sqrt{5}$ & 57 & 0.0355 & 0.1274 \\
$\sqrt{6}$ & 81 & 0.0285 & 0.0961 \\
$\sqrt{8}$ & 93 & 0.0242 & 0.1025 \\
\end{tabular}
\label{table2}
\end{table}
}

\newpage

\centerline{\bf Figure Captions}
\vspace{.2in}

\begin{itemize}
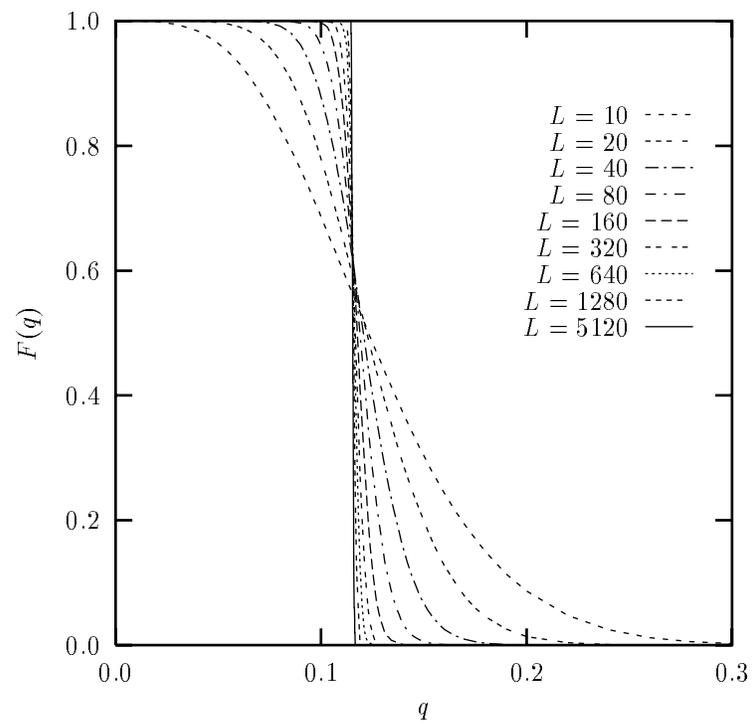
\item[Fig.~1.] Variation of the percolating fraction $F(q)$ is plotted
against the probability $q$, for lattice sizes $L$=10, 20, 40, 80, 160,
320, 640, 1280, and
5120, and with radius $r$=1 (lattice unit), for a two
dimensional square lattice (plot for $L=2560$ is not shown for the sake of
clarity). Data are averaged over 100000 realizations
for $L$=10, 50000 for $L$=20, 20000 for $L$=40, 10000 for $L$=80,
5000 for $L$=160, 2000 for $L$=320, 1000 for $L$=640,
500 for $L$=1280, and 50 for $L$=5120.
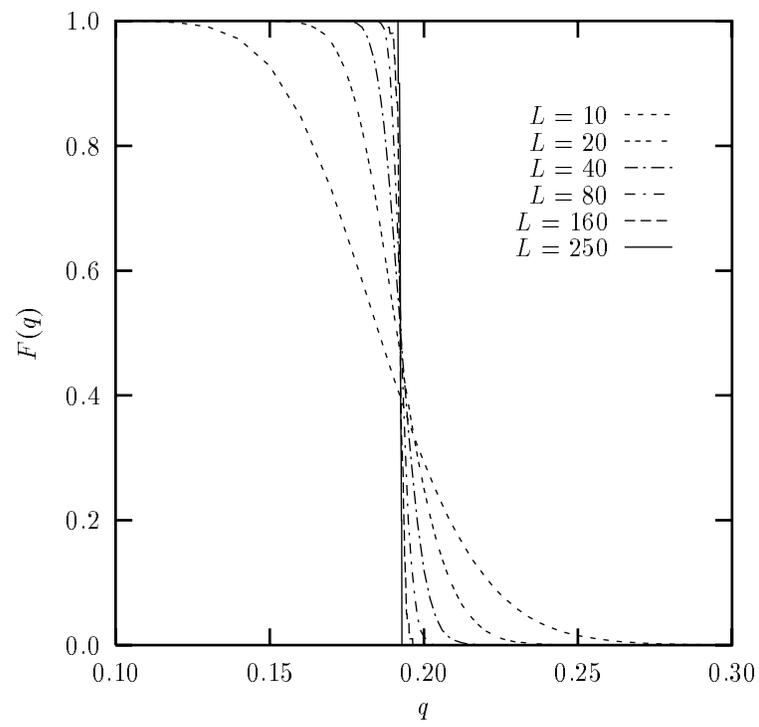
\item[Fig.~2.] $F(q)$ vs. $q$ plot for three dimensional simple cubic
lattice, for $L$=10, 20, 40, 80, 160, and 250, with $r$=1. Realizations
taken are 100000 for $L$=10, 30000 for $L$=20, 10000 for $L$=40, 2000 for
$L$=80, 200 for $L$=160, and 10 for $L$=250.
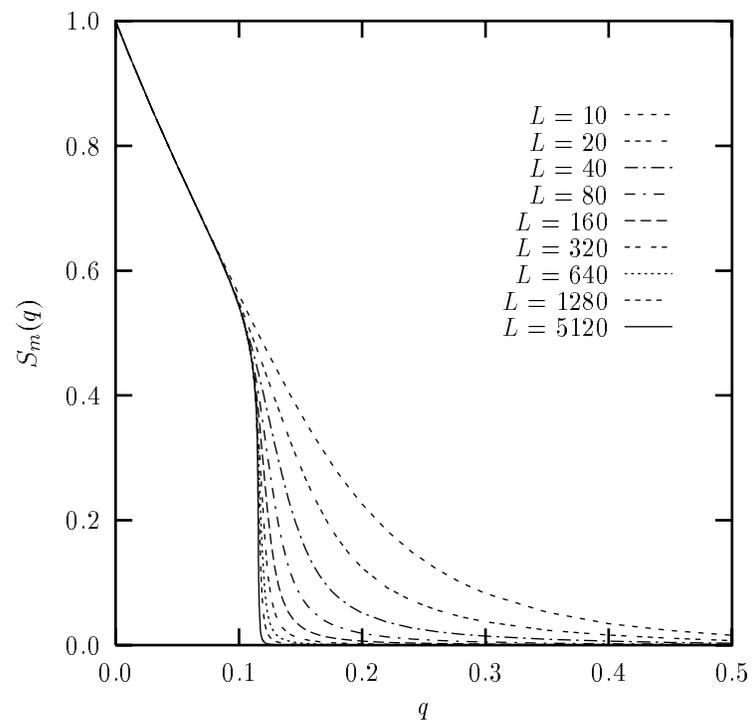
\item[Fig.~3.] Variation of the maximum size $S_m(q)$ of the cluster
is plotted against $q$, for the same $L$ and $r$ values as in Fig.~1.
\end{itemize}

\end{multicols}

\begin{thebibliography}{99}
\bibitem{SA} D. Stauffer and A. Aharony, {\it Introduction to Percolation
Theory} (Taylor \& Francis, London, 1991).
\bibitem{JWE} J. W. Essam, Rep. Prog. Phys. {\bf 43}, 843 (1980).
\bibitem{HK} H. Kesten, {\it Percolation Theory for Mathematicians}
(Birkhauser, Boston, 1982).
\bibitem{GM} S. Galam and A. Mauger, Phys. Rev. E {\bf 53}, 2177 (1996).
\bibitem{JSW} J. S. Wang et al., Physica A {\bf 167}, 565 (1990).
\bibitem{SK} S. Kirkpatrick, Rev. Mod. Phys. {\bf 45}, 574 (1973).
\bibitem{AMAHK} J. Adler, Y. Meir, A. Aharony, A. B. Harris, and L.
Klein, J. Stat. {\bf 58}, 511 (1990).
\bibitem{FHS} S. Feng, B. I. Halperin and P. Sen, Phys. Rev. B {\bf 35},
197 (1987).
\bibitem{EG} E. N. Gilbert, SIAM J. Appl. Math. {\bf 9}, 533 (1961).
\bibitem{FK} S. J. Fraser and R. Kapral, J. Chem. Phys. {\bf 85}, 5682
(1986).
\bibitem{MS} M. Sahimi, {\it Applications of Percolation Theory}
(Taylor \& Francis, London, 1994).
\bibitem{BK} P. J. M. Bastiaansen and H. J. F. Knops, preprint (1996).
\bibitem{OSP} S. B. Ogale, R. Shreekala, S. I. Patil, B. Hannoyer, F.
Pettit and G. Marest, to be published.
\bibitem{FR} There is a strong belief that magnetic polarons do play a
significant role in colossal magnetoresistance properties of these oxide
perovskite systems \cite{MLS,KYT}. We subscribe to this belief. 
\bibitem{MLS} A. J. Millis, P. B. Littlewood, and B. I. Shraiman, Phys.
Rev. Lett. {\bf 74}, 5144 (1995).
\bibitem{KYT} T. Kasuya, A. Yanase and T. Takeda, Sol. State Comm.
{\bf 8}, 1551 (1970).
\bibitem{AR1} There exist other such equivalences for objects
or obstacles with more general symmetries. 
\bibitem{AR2} All numerical calculations have been
carried out on a Silicon Graphics Workstation with a 64--bit R8000
processor and 64 Mb RAM.
\bibitem{AR3} R. E. Amritkar and Manojit Roy, to be published.
\bibitem{AR4} We emphasize that our model is essentially a classical
model and therefore corrections due to quantum effects are expected.
\bibitem{AR5} One may compare our model with continuum percolation problem
\cite{SA,FHS}. However, it may be noted that for the physical example 
considered here, ours is the correct model to study it, whereas continuum
model is not physically relevant.
\end{thebibliography}
\end{document}